\documentstyle[12pt]{article}

\newcommand{\be}{\begin{equation}}
\newcommand{\ee}{\end{equation}}
\newcommand{\bea}{\begin{eqnarray}}
\newcommand{\eea}{\end{eqnarray}}
\newcommand{\beaa}{\begin{eqnarray*}}
\newcommand{\eeaa}{\end{eqnarray*}}

\newcommand{\BB}{{{\rm I} \kern -2pt \rlap {\rm B} \kern +8pt}}

\advance\textheight by \topskip
\makeatletter

\@addtoreset{equation}{section}

\def\section{\@startsection {section}{1}{\z@}{-3.5ex plus -1ex minus
 -.2ex}{2.3ex plus .2ex}{\large\bf\centering}}
\def\subsection{\@startsection{subsection}{2}{\z@}{-3.25ex plus%
 -1ex minus -.2ex}{1.5ex plus .2ex}{\bf}}
\def\subsubsection{\@startsection{subsubsection}{3}{\z@}{-3.25ex plus%
 -1ex minus -.2ex}{1.5ex plus .2ex}{\sl}}
\makeatother

\begin{document}

\baselineskip 18pt
\parindent 12pt
\parskip 10pt

\begin{titlepage}

\begin{center}
{\Large {\bf Superfield Lax formalism of supersymmetric sigma
model on symmetric spaces }}\\\vspace{1.5in} {\large
U. Saleem\footnote{%
usman\_physics@yahoo.com} and M. Hassan \footnote{%
mhassan@physics.pu.edu.pk} }\vspace{0.15in}

{\small{\it Department of Physics,\\ University of the Punjab,\\
Quaid-e-Azam Campus,\\Lahore-54590, Pakistan.}}
\end{center}

\vspace{1cm}
\begin{abstract}
We present a superfield Lax formalism of superspace sigma model
based on the target space ${\cal G}/{\cal H}$ and show that a
one-parameter family of flat superfield connections exists if the
target space ${\cal G}/{\cal H}$ is a symmetric space. The
formalism has been related to the existences of an infinite family
of local and non-local superfield conserved quantities. Few
examples have been given to illustrate the results.
\end{abstract}
\end{titlepage}
\section{Introduction}
In recent years, some investigations have been made to study the
infinite number of conserved quantities of symmetric space sigma
models, their Poisson bracket algebra and the quantum conservation
of these quantities \cite{J2}-\cite{J1}. The supersymmetric
extension of sigma models based on symmetric spaces ${\cal
G}/{\cal H}$ have also been studied quite recently, and it has
been shown that there exist two classes of local conservation
laws; one class of conservation laws corresponds to cohomology of
the target manifold and the second class corresponds to higher
spin generalizations of the energy momentum tensor \cite{J3}. The
investigation of these integrable models has applications in
recent advances in superstring theories on AdS backgrounds
\cite{ads1}-\cite{ads9}. The model studies in these investigations
are in many ways related to the symmetric space sigma model. In
all these studies, a formal analysis of integrability of
supersymmetric sigma models on symmetric space has not been
carried out so far, which we intend to present in this work.

In this paper we will extend certain results related to the
integrability of the supersymmetric sigma models based on ${\cal
G}/{\cal H}$. We generalize earlier results of Eichenherr and
Forger \cite{EIC} and show that the supersymmetric sigma model
with target space a homogeneous space ${\cal G}/{\cal H}$ admits a
one-parameter family of flat superfield currents if ${\cal
G}/{\cal H}$ is a symmetric space. Our main result is to find a
superfield Lax formalism in terms of a one-parameter family of
flat superfield currents of supersymmetric sigma model on
symmetric space and relate it to the infinitely many local and
non-local superfield conserved quantities of the model. We
illustrate our results by giving some explicit examples of sigma
models on complex Grassmannians $U(m+n)/U(m)\times U(n)$ and
principal chiral models
for which the corresponding symmetric spaces are ${\cal G}\times {\cal G}/%
{\cal G}$.

The paper is organized as follows. In the section \ref{section2},
we give a general theory of ${\cal G}/{\cal H}$ sigma model in
superspace. Section \ref{section3} contains the Lax formalism of
the model in terms of one-parameter family of flat superfield
connections, and it has been shown that a Lax formalism is
admissible when ${\cal G}/{\cal H}$ is a symmetric space. In
section \ref{section4}, we investigate the existence of local and
non-local conserved quantities of the model and compare our
results with the results obtained earlier. Section \ref{section5}
gives explanation of our investigations with some explicit
examples. Section \ref{section6} contains our concluding remarks.

\section{The ${\cal G}/{\cal H}$ sigma model in
superspace}\label{section2}

We begin by defining a compact symmetric space and formulate
supersymmetric non-linear sigma model on a symmetric space ${\cal
G}/{\cal H}$. For the general structure of the model, we follow
the treatment of the model adopted for the bosonic model
\cite{EIC}-\cite{john2} and the supersymmetric model \cite{J3}.

A symmetric space is defined as follows. Let ${\cal G}$ be a
compact Lie group with Lie algebra ${\bf g}$ and let ${\cal H}$ be
its subgroup with a Lie
algebra ${\bf h}$. Let $\sigma $ be a linear automorphism $\sigma :{\bf g}%
\rightarrow {\bf g}$ such that $\sigma ^{2}=1$. This means that
$\sigma $ has eigenvalues $\pm 1$ and it splits the algebra ${\bf
g}$ into orthogonal eigenspaces corresponding to these
eigenvalues. This automorphism is called an involutive
automorphism. This causes the canonical decomposition of ${\bf g}$
as follows:
\begin{equation}
{\bf g}={\bf h+k,}  \label{decom}
\end{equation}
where ${\bf h}$  and ${\bf k}$ are the $(+1)$ and $(-1)$
eigenspaces of $\sigma $ ($Z_{2}$ grading of ${\bf g}$) and the
action of $\sigma $ on vectors of ${\bf g,h}$ and ${\bf k}$ is as
follows:
\begin{eqnarray*}
\left[ \sigma (X),\sigma (Y)\right] &=&\sigma \left[ X,Y\right]\,\,\mbox{%
for }X,Y\in {\bf g,} \\
\sigma (X) &=&X\mbox{ for }X\in {\bf h,} \\
\sigma (X) &=&-X\mbox{ for }X\in {\bf k.}
\end{eqnarray*}
It is clear from the above that ${\bf h}$ is a subalgebra but
${\bf k}$ is not. In fact, ${\bf k}$ is the orthogonal complement
of ${\bf h}$ in ${\bf g}$. The following Lie brackets hold:
\begin{equation}
\lbrack {\bf h},{\bf h}]\subset {\bf h},\ \ [{\bf k},{\bf k}]\subset {\bf %
k,\,\,}[{\bf h},{\bf k}]\subset {\bf k}.  \label{symmetric}
\end{equation}
The algebra ${\bf h}$ satisfying the relation (\ref{symmetric}) is
called a symmetric space subalgebra. The coset homogeneous space
${\cal G}/{\cal H}$ with involutive automorphism $\sigma $ and
admitting a canonical decomposition (\ref{decom}) obeying relation
(\ref{symmetric}), is called a symmetric space
\cite{space1,space2}.

In order to formulate a supersymmetric sigma model on ${\cal
G}/{\cal H}$ in $(1+1)$-dimensions\footnote{Our notation
conventions are as follows. The two-dimensional Minkowski metric
is $\eta _{\mu \nu }=$diag$(+1,-1)$ and the orthonormal and
light-cone coordinates are related by $x^{\pm}=\frac{1}{2}(x_{0}
\pm x_{1})$ and $\partial_{\pm}=\frac{1}{2}(\partial_{0} \pm
\partial_{1})$. Under a Lorentz transformation $x^{\pm }$ and $\partial _{\pm }$
transform as $x^{\pm }\longmapsto e^{\mp \Lambda }x^{\pm }$ and
$\partial _{\pm }\longmapsto e^{\mp \Lambda }\partial _{\pm }$,
where $\Lambda $ is the rapidity of the Lorentz boost.} , we
define the superfield $Q(x^{\pm },\theta ^{\pm })$ as a function
of space-time coordinates $x^{\pm}$ and anti-commuting coordinates
$\theta^{\pm}$ and taking values in ${\cal G}/{\cal H}$ and it is
lifted (locally) to the superfield $G(x^{\pm },\theta ^{\pm })$
taking values in ${\cal G}$, with a natural equivalence
\[
G_{2}(x^{\pm },\theta ^{\pm })\sim G_{1}(x^{\pm },\theta ^{\pm }),
\]
such that there exists a superfield $H(x^{\pm },\theta ^{\pm })\in
{\cal H}$ , such that both are related by a gauge transformation
\begin{equation}
G_{2}(x^{\pm },\theta ^{\pm })=G_{1}(x^{\pm },\theta ^{\pm
})H(x^{\pm },\theta ^{\pm }).  \label{gau}
\end{equation}
For the gauge invariant quantities such as the superfield
$Q(x^{\pm },\theta ^{\pm })$ ordinary derivatives are relevant,
while on gauge covariant quantities such as the superfield
$G(x^{\pm },\theta ^{\pm })$, the ordinary derivatives have to be
replaced by covariant derivatives. We define the gauge covariant
derivative in superspace acting on the superfield $G(x^{\pm
},\theta ^{\pm })$ by
\begin{equation}
{\cal D}_{\pm }G=D_{\pm }G-iG{\cal A}_{\pm },  \label{co}
\end{equation}
where $iG{\cal A}_{\pm }\equiv \pi (D_{\pm }G)$ is the vertical part of $%
D_{\pm }G$ and ${\cal D}_{\pm }G\equiv (1-\pi )(D_{\pm }G)$ is the
horizontal part of $D_{\pm }G$ \footnote{The super derivatives
$D_{\pm}$ are defined as $D_{\pm }=\frac{\partial }{\partial
\theta ^{\pm }}-i\theta ^{\pm }\partial _{\pm },
D_{\pm}^2=-i\partial_{\pm}, \{D_{+},D_{-}\}=0$, where \{,\} is an
anti-commutator. The supersymmetry generators are
$Q_{\pm}=\partial_{\theta^{\pm}}+i\partial_{\pm}$, obeying
$Q^{2}_{\pm}=i\partial_{\pm}$.} \cite{J3}. The gauge invariant
conserved superfield currents are
\begin{equation}
J_{\pm }\equiv i\alpha {\cal D}_{\pm }GG^{-1}=i\alpha (1-\pi
)D_{\pm }GG^{-1},  \label{incurrent}
\end{equation}
and gauge-covariant conserved superfield currents
\begin{equation}
K_{\pm }\equiv -i\alpha G^{-1}{\cal D}_{\pm }G=-i\alpha (1-\pi
)G^{-1}D_{\pm }G,  \label{cocurrent}
\end{equation}
where $\alpha $ is some real constant introduced for later
convenience. Under the gauge transformation the superfield
$G(x^{\pm },\theta ^{\pm })$ transforms as $G(x^{\pm },\theta
^{\pm })\rightarrow G(x^{\pm },\theta ^{\pm })H(x^{\pm },\theta
^{\pm })$ and the corresponding gauge invariant and covariant
superfield currents transform as
\[
J_{\pm }\rightarrow J_{\pm
}\,,\,\,\,\,\,\,\,\,\,\,\,\,\,\,\,\,\,\,\,\,\,\,\,\,\,\,\,\,\,\,\,\,\,\,\,\,%
\,\,\,\,\,\,\,\,\,\,\,\,\,\,\,\,\,\,\,\,K_{\pm }\rightarrow
H^{-1}K_{\pm }H.
\]
We define the action of covariant derivative in superspace on the
gauge covariant superfield currents $K_{\pm }$ by
\[
{\cal D}_{\pm }K_{\mp }=D_{\pm }K_{\mp }+i\left\{ {\cal A}_{\pm
}\,,K_{\mp }\right\} .
\]
The lagrangian for the ${\cal G}/{\cal H}$ sigma model in
superspace is \cite{J3}
\begin{equation}
{\cal L}_{{\cal G}/{\cal H}}\equiv{\frac{1}{2}}\mbox{Tr}\left( {\cal D}_{+}G^{-1}%
{\cal D}_{-}G\right) ={\frac{1}{2}}\mbox{Tr}\left(
D_{+}Q^{-1}D_{-}Q\right) . \label{action}
\end{equation}
The superspace equations of motion obtained from the lagrangian
can be expressed as
\begin{eqnarray}
D_{-}J_{+}-D_{+}J_{-} &=&0,  \label{con1} \\
{\cal D}_{-}K_{+}-{\cal D}_{+}K_{-} &=&0.  \label{con2}
\end{eqnarray}
For the gauge superfields ${\cal A}_{\pm }$, we can have the
following equation for the given homogeneous space ${\cal G}/{\cal
H}$
\begin{equation}
D_{-}{\cal A}_{+}+D_{+}{\cal A}_{-}=i\pi \left\{ G^{-1}{\cal D}_{+}G+i{\cal A%
}_{+},G^{-1}{\cal D}_{-}G+i{\cal A}_{-}\right\}, \label{ss}
\end{equation}
the curvature form in terms of the superfields ${\cal A}_{\pm }$
is
\begin{equation}
{\cal F}_{\,-+}\equiv D_{-}{\cal A}_{+}+D_{+}{\cal A}_{-}+i\left\{ {\cal A}%
_{+},{\cal A}_{-}\right\} =i\pi \left\{ G^{-1}{\cal D}_{+}G,G^{-1}{\cal D}%
_{-}G\right\} .  \label{cur}
\end{equation}
In the above expression ${\cal F}_{\,-+}$ represents the curvature
form of the gauge superfields ${\cal A}_{\pm }$. Our aim in this
paper is to develop a superfield Lax formalism of the model in
terms of a one parameter family of flat superfield connections. In
the sections which follow, we show that the existence of flat
superfield connections is admissible when the target space $\cal
G/\cal H$ of the model is a symmetric space. The superfield Lax
formalism is then used to derive infinitely many local and
non-local superfield conserved quantities.

\section{Superfield Lax formalism}\label{section3}

The superfield Lax formalism of the symmetric space sigma model in
superspace can be obtained by defining a one-parameter family of
transformations on superfields of the model. A one-parameter
family of transformations on the superfiels is defined in terms of
matrix superfields ${\cal U}^{(\gamma )}$ which obey the following
set of linear differential equations
\begin{eqnarray}
D_{+}{\cal U}^{(\gamma )} &\equiv &i(1-\gamma ^{-1}){\cal
U}^{(\gamma )}\,J_{+}=-\alpha (1-\gamma ^{-1}){\cal U}^{(\gamma
)}{\cal D}_{+}GG^{-1},
\nonumber \\
D_{-}{\cal U}^{(\gamma )} &\equiv &i(1-\gamma ){\cal U}^{(\gamma
)}\,J_{-}=-\alpha (1-\gamma ){\cal U}^{(\gamma )}{\cal
D}_{-}GG^{-1}, \label{one}
\end{eqnarray}
where the matrix superfield ${\cal U}\in {\cal G}$. The model
retains the zero-curvature representation if the compatibility
condition of the linear system (\ref{one}) becomes equivalent to
the following equation
\begin{equation}
(1-\gamma ^{-1})D_{-}J_{+}+(1-\gamma )D_{+}\,J_{-}+i(1-\gamma
^{-1})(1-\gamma )\left\{ J_{+},J_{-}\right\} =0.  \label{zero}
\end{equation}
Using equation (\ref{con1}) the compatibility condition
(\ref{zero}) reduces to $\gamma $-independent equation
\begin{equation}
D_{-}J_{+}+D_{+}\,J_{-}+2i\left\{ J_{+},J_{-}\right\} =0,
\label{zero1}
\end{equation}
which is the zero-curvature condition for the superfield currents $J_{\pm }$%
. If we look at the compatibility condition of the linear system (\ref{one}%
), we arrive at the following $\gamma$-independent equation
\begin{equation}
D_{-}J_{+}+D_{+}\,J_{-}+2i\left\{ J_{+},J_{-}\right\} =-i\alpha
(2\alpha -1-\pi )G\left\{ G^{-1}{\cal D}_{+}G,G^{-1}{\cal
D}_{-}G\right\} G^{-1}, \label{zero2}
\end{equation}
which does not indicate that the superfield current $J_{\pm }$ is
flat. In order to formulate a theory which gives rise to flat
superfield currents, we look at the extra terms appearing on the
right hand side of the equation (\ref{zero2}) and look at the
constraints which appear in the geometric structure of the target
space when these extra terms are set equal to zero. To achieve
this we discuss two different cases.

In the first case we assume that $[{\bf k},{\bf k}]\subset {\bf h}$%
, which implies that
\[ \left( 1-\pi \right) \left\{ G^{-1}{\cal
D}_{+}G,G^{-1}{\cal D}_{-}G\right\} =0,
\]
the equation (\ref{zero2}) reduces to (\ref{zero1}), if we choose $\alpha =1$%
. The matrix superfield ${\cal U}^{(\gamma )}$ generates a
one-parameter family of transformations on the solutions of the
equations of motion. The superfield $G(x^{\pm },\theta ^{\pm })$
transforms as
\begin{equation}
G(x^{\pm },\theta ^{\pm })\rightarrow G^{(\gamma )}(x^{\pm },\theta ^{\pm })=%
{\cal U}^{(\gamma )}G(x^{\pm },\theta ^{\pm }),  \label{tran1}
\end{equation}
where $G^{(1)}(x^{\pm },\theta ^{\pm })=G(x^{\pm },\theta ^{\pm
})$. The action of the derivatives $D_{\pm }$ on the superfield
$G^{(\gamma )}(x^{\pm },\theta ^{\pm })$ will be
\[
D_{\pm }G^{(\gamma )}=\gamma ^{\mp1}{\cal U}^{(\gamma )}{\cal
D}_{\pm }G+iG^{(\gamma )}{\cal A}_{\pm },
\]
where ${\cal U}^{(\gamma )}{\cal D}_{\pm }G$ is the horizontal part and $%
iG^{(\gamma )}{\cal A}_{\pm }$ is the vertical part of $D_{\pm
}G$. This implies
\begin{equation}
{\cal D}_{\pm }^{(\gamma )}G^{(\gamma )}=\gamma ^{\mp1}{\cal U}^{(\gamma )}%
{\cal D}_{\pm }G,\,\ \ \,{\cal A}_{\pm }^{(\gamma )}={\cal A}_{\pm
}. \label{tran2}
\end{equation}
where ${\cal D}_{\pm }^{(\gamma )}G^{(\gamma )}=D_{\pm }G^{(\gamma
)}-iG^{(\gamma )}\,{\cal A}_{\pm }^{(\gamma )}.\,$\thinspace
\thinspace Thus it can be seen that the Lagrangian (\ref{action})
is invariant under the transformation (\ref{tran1}).

The second case is when $[{\bf k},{\bf k}]\subset {\bf k,}$ which
is equivalent to say that
\[
\pi \left\{ G^{-1}{\cal D}_{+}G,G^{-1}{\cal D}_{-}G\right\} =0.\
\]
The equation (\ref{zero2}) reduces to (\ref{zero1}), if we choose $\alpha =%
\frac{1}{2}$. For this case, consider the following one-parameter
family of differential equations
\begin{eqnarray}
{\cal D}_{+}{\cal V}^{(\gamma )} &\equiv &i(1-\gamma ^{-1}){\cal
V}^{(\gamma )}K_{+}={\frac{1}{2}}(1-\gamma ^{-1}){\cal V}^{(\gamma
)}G^{-1}{\cal D}_{+}G,
\nonumber \\
{\cal D}_{-}{\cal V}^{(\gamma )} &\equiv &i(1-\gamma ){\cal
V}^{(\gamma )}K_{-}={\frac{1}{2}}(1-\gamma ){\cal V}^{(\gamma
)}G^{-1}{\cal D}_{-}G. \label{one1}
\end{eqnarray}
where $K_{\pm }(x^{\pm },\theta ^{\pm })$ are the components of
the gauge covariant superfields. The ${\cal G}$-valued matrix
superfield ${\cal V}^{(\gamma )}$ transforms under the gauge
transformation as
\[
{\cal V}^{(\gamma )}\rightarrow {\cal V}^{(\gamma )}H(x^{\pm
},\theta ^{\pm }).
\]
By using this gauge transformation, the system (\ref{one1}) can
also be expressed as
\begin{eqnarray}
D_{+}{\cal V}^{(\gamma )} &\equiv &i(1-\gamma ^{-1}){\cal
V}^{(\gamma
)}K_{+}+i{\cal V}^{(\gamma )}{\cal A}_{+}={\frac{1}{2}}(1-\gamma ^{-1}){\cal %
V}^{(\gamma )}G^{-1}{\cal D}_{+}G+i{\cal V}^{(\gamma )}{\cal
A}_{+},
\nonumber \\
D_{-}{\cal V}^{(\gamma )} &\equiv &i(1-\gamma ){\cal V}^{(\gamma )}K_{-}+i%
{\cal V}^{(\gamma )}{\cal A}_{-}={\frac{1}{2}}(1-\gamma ){\cal
V}^{(\gamma )}G^{-1}{\cal D}_{-}G+i{\cal V}^{(\gamma )}{\cal
A}_{-}.  \label{one2}
\end{eqnarray}
The compatibility condition of the linear system (\ref{one2}) is
\begin{eqnarray}
&& D_{-}\left( \left( 1-\gamma ^{-1}\right) K_{+}+{\cal
A}_{+}\right) +D_{+}\left( \left( 1-\gamma \right) K_{-}+{\cal
A}_{-}\right)  +i\left\{ (1-\gamma ^{-1})K_{+}+{\cal
A}_{+},(1-\gamma )K_{-}+{\cal A}_{-}\right\}
\nonumber \\
&&=(1-\gamma ^{-1}){\cal D}_{-}K_{+}+(1-\gamma ){\cal
D}_{+}K_{-}+i(1-\gamma ^{-1})(1-\gamma )\left\{
K_{+},K_{-}\right\}+{\cal F}_{-+}.\label{A}
\end{eqnarray}
For this case, the equation (\ref{cur}) implies that curvature form ${\cal F}%
_{-+\,}$ vanishes so that the compatibility condition (\ref{A})
reduces to the following $\gamma $-independent equation
\[
{\cal D}_{-}K_{+}+{\cal D}_{+}K_{-}+2i\left\{ K_{+},K_{-}\right\} \equiv -%
\frac{1}{2}{\cal F}_{-+\,}=0.
\]
The matrix superfields ${\cal U}^{(\gamma )}$ and ${\cal
V}^{(\gamma )}$ generate a one-parameter family of transformations
on the solutions of the superfield equations which gives rise to a
one-parameter family of flat superfield currents. This particular
transformation is given by
\begin{equation}
G(x^{\pm },\theta ^{\pm })\rightarrow G^{(\gamma )}(x^{\pm },\theta ^{\pm })=%
{\cal U}^{(\gamma )}G(x^{\pm },\theta ^{\pm }){\cal V}^{(\gamma )-1}{\cal V}%
^{(1)}.  \label{gtran1}
\end{equation}
The action of derivatives $D_{\pm }$ and ${\cal D}_{\pm }$ on the
superfields $G^{(\gamma )}(x^{\pm },\theta ^{\pm })$ is
\begin{eqnarray*}
D_{\pm }G^{(\gamma )} &=&\gamma ^{\mp1}{\cal U}^{(\gamma )}{\cal D}_{\pm }G%
{\cal V}^{(\gamma )-1}{\cal V}^{(1)}+iG^{(\gamma )}{\cal A}_{\pm }, \\
{\cal D}_{\pm }^{(\gamma )}G^{(\gamma )} &=&\gamma ^{\mp1}{\cal U}^{(\gamma )}%
{\cal D}_{\pm }G{\cal V}^{(\gamma )-1}{\cal V}^{(1)},\,\,\,\,\,\,\,\,\,\,\,%
\,\,\,\,\,\,\,\,\,\,{\cal A}_{\pm }={\cal A}_{\pm }^{(\gamma )},
\end{eqnarray*}
where ${\cal D}_{\pm }^{(\gamma )}G^{(\gamma )}=D_{\pm }G^{(\gamma
)}-iG^{(\gamma )}{\cal A}_{\pm }^{(\gamma )}$. The Lagrangian (\ref{action}%
) is invariant under transformation (\ref{gtran1}).

In the first case, where we have $[{\bf k},{\bf k}]\subset {\bf
h}$, the linear map $\sigma:{\bf g}\rightarrow {\bf g}$ is an
isometric Lie algebra automorphism and can be lifted to an
isometric Lie group automorphism, which is always true if $\cal
{G}$ is simply connected. The coset space $\cal {G}/\cal {H}$ is
then a symmetric space \cite{space1,space2}. In the second case,
where we have $[{\bf k},{\bf k}]\subset {\bf k}$, the coset space
$\cal {G}/\cal {H}$ is canonically isomorphic to the connected
normal Lie subgroup $\cal {K}$ in the Lie group $\cal {G}$
generated by ${\bf k}$ in ${\bf g}$, and the action of $\cal {H}$
on $\cal {K}$ by the Lie algebra automorphism $\sigma$ can be
lifted to an action of $\cal {H}$ on $\cal {K}$ by the Lie group
automorphism which is always the case if $\cal {K}$ is simply
connected. For a special case when $\cal {K}$ is canonically
isomorphic to the symmetric space
\[\cal {K}\times\cal
{K}/\bigtriangleup\cal {K},
\]
we get the principal chiral model discussed in the section
\ref{section5}. In summary: we have observed that for the first
case we have to take $\cal {G}$ to be simply connected and for the
second case $\cal {K}$ must be compact. This shows that the
supersymmetric sigma model based on ${\cal G}/{\cal H}$ admits a
Lax formalism and zero-curvature representation presented in
(\ref{one}) and (\ref{zero2}) respectively if ${\cal G}/{\cal H}$
is a symmetric space. This is essentially a supersymmetric
generalization of earlier work of Eichenerr and Forger \cite{EIC}.
In what follows, we explicitly write the Lax pair of the model and
derive an infinite family of superfield conserved quantities.

In both cases, a one-parameter family of flat superfield currents
is given by the following transformation rule
\begin{eqnarray*}
J_{+} &\mapsto &J_{+}^{(\gamma )}\,=\,\gamma ^{-1}{\cal U}{^{(\gamma )}}J_{+}%
{\cal U}{^{(\gamma )}}^{-1}\;, \\
J_{-} &\mapsto &J_{-}^{(\gamma )}\,=\,\gamma \,\,{\cal U}{^{(\gamma )}}J_{-}%
{\cal U}{^{(\gamma )}}^{-1}\;.
\end{eqnarray*}
These superfield currents are conserved in superspace for any value of $%
\gamma $: $D_{+}J_{-}^{\,(\gamma )}-D_{-}J_{+}^{(\gamma )}=0$. The
associated linear system of the supersymmetric sigma model on
symmetric space can be written as
\begin{equation}
D_{\pm }{\cal U}(t,x,\theta ;\lambda )={\cal U}(t,x,\theta ;\lambda )\,{\cal %
P}_{\pm }^{(\lambda )},  \label{Lax1}
\end{equation}
where the odd superfields ${\cal P}_{\pm }^{(\lambda )}$ are given
by
\[
{\cal P}_{\pm }^{(\lambda )}=\mp \frac{2i\lambda }{(1\mp \lambda
)}J_{\pm }.
\]
The parameter $\lambda $ is the spectral parameter and is related
to the parameter $\gamma $ by $\lambda =\frac{1-\gamma }{1+\gamma
}$. The compatibility condition of the linear system (\ref{Lax1})
reduces to a fermionic zero-curvature condition for the odd
superfields ${\cal P}_{\pm }^{(\lambda )}$ as
\[
\left\{ D_{+}-{\cal P}_{+}^{(\lambda )},D_{-}-{\cal
P}_{-}^{(\lambda )}\right\} \equiv D_{-}{\cal P}_{+}^{(\lambda
)}+D_{+}{\cal P}_{-}^{(\lambda )}+\left\{ {\cal P}_{+}^{(\lambda
)},{\cal P}_{-}^{(\lambda )}\right\} =0.
\]
Now we can define the superspace Grassmann odd operators ${\cal
L}_{\pm
}^{(\lambda )}$%
\[
{\cal L}_{\pm }^{(\lambda )}=D_{\pm }-{\cal P}_{\pm }^{(\lambda
)},
\]
obeying the (Lax) equations in superspace
\[
D_{\mp }{\cal L}_{\pm }^{(\lambda )}=\left\{ {\cal P}_{\mp }^{(\lambda )},%
{\cal L}_{\pm }^{(\lambda )}\right\} .
\]
By applying $D_{\pm }$ on (\ref{Lax1}), one gets a linear system
in terms of
even superfields $\tilde{{\cal P}}_{\pm }^{(\lambda )}$%
\begin{equation}
\partial _{\pm }{\cal \,U}(t,x,\theta ;\lambda )={\cal U}(t,x,\theta
;\lambda )\tilde{{\cal P}}_{\pm }^{(\lambda )},  \label{even}
\end{equation}
$\;$where the even superfields $\tilde{{\cal P}}_{\pm }^{(\lambda
)}\,$ are given by
\[
\tilde{{\cal P}}_{\pm }^{(\lambda )}=\left\{\pm \left(
\frac{2\lambda }{1\mp
\lambda }\right) D_{\pm }J_{\pm }-i\left( \frac{2\lambda }{1\mp \lambda }%
\right) ^{2}J_{\pm }^{2}\right\} .
\]
The compatibility condition of the linear system (\ref{even}) now
reduces to a bosonic zero-curvature condition for the even superfields $\tilde{{\cal P}}%
_{\pm }^{(\lambda )}$
\[
\left[ \partial _{+}-\tilde{{\cal P}}_{+}^{(\lambda )},\partial _{-}-\tilde{%
{\cal P}}_{-}^{(\lambda )}\right] \equiv \partial _{-}\,\,\tilde{{\cal P}}%
_{+}^{(\lambda )}-\partial _{+}\,\tilde{{\cal P}}_{-}^{(\lambda
)}+\left[ \tilde{{\cal P}}_{+}^{(\lambda )},\tilde{{\cal
P}}_{-}^{(\lambda )}\right] =0.
\]
The superspace Grassmann even Lax operators $\tilde{{\cal L}}_{\pm
}^{(\lambda )}$%
\[
\tilde{{\cal L}}_{\pm }^{(\lambda )}=\partial _{\pm }-\tilde{{\cal
P}}_{\pm }^{(\lambda )},
\]
obey the following equation
\[
\partial _{\mp }\tilde{{\cal L}}_{\pm }^{(\lambda )}=\left[ \tilde{{\cal P}}%
_{\mp }^{(\lambda )},\tilde{{\cal L}}_{\pm }^{(\lambda )}\right] .
\]
The linear system (\ref{even}) can be re-expressed in terms of
space-time coordinates by
\begin{equation}
\;\partial _{0}{\cal \,U}(t,x,\theta ;\lambda )=\,{\cal
U}(t,x,\theta ;\lambda )\tilde{{\cal P}}_{0}^{(\lambda
)},\,\,\,\,\,\,\,\,\,\,\,\,\,\,\,\,\,\,\,\,\partial _{1}{\cal \,U}%
(t,x,\theta;\lambda )=\,{\cal U}(t,x,\theta ;\lambda )\tilde{{\cal
P}}_{1}^{(\lambda )},  \label{even2}
\end{equation}
with the superfields $\tilde{{\cal P}}_{0}^{(\lambda )}$and $\tilde{{\cal P}}%
_{1}^{(\lambda )}$ defined by
\begin{eqnarray*}
\tilde{{\cal P}}_{0}^{(\lambda )} &=&\frac{1}{2}\left\{ \left( \frac{%
2\lambda }{1-\lambda }\right) D_{+}J_{+}-i\left( \frac{2\lambda }{1-\lambda }%
\right) ^{2}J_{+}^{2}-\left( \frac{2\lambda }{1+\lambda }\right)
D_{-}J_{-}-i\left( \frac{2\lambda }{1+\lambda }\right)
^{2}J_{-}^{2}\right\}
, \\
\tilde{{\cal P}}_{1}^{(\lambda )} &=&\frac{1}{2}\left\{ \left( \frac{%
2\lambda }{1-\lambda }\right) D_{+}J_{+}-i\left( \frac{2\lambda }{1-\lambda }%
\right) ^{2}J_{+}^{2}+\left( \frac{2\lambda }{1+\lambda }\right)
D_{-}J_{-}+i\left( \frac{2\lambda }{1+\lambda }\right)
^{2}J_{-}^{2}\right\} .
\end{eqnarray*}This is the bosonic superfield Lax pair of the
model. The compatibility condition of the linear system
(\ref{even2}) therefore becomes
\[
\left[ \partial _{0}-\tilde{{\cal P}}_{0}^{(\lambda )},\partial _{1}-\tilde{%
{\cal P}}_{1}^{(\lambda )}\right] \equiv \partial _{1}\,\,\tilde{{\cal P}}%
_{0}^{(\lambda )}-\partial _{0}\,\tilde{{\cal P}}_{1}^{(\lambda
)}+\left[ \tilde{{\cal P}}_{0}^{(\lambda )},\tilde{{\cal
P}}_{1}^{(\lambda )}\right] =0.
\]
We now define the theory on the spatial interval $\left[
-a,a\right] $ and the
superfields $\tilde{{\cal P}}_{0}^{(\lambda )}$ and $\tilde{{\cal P}}%
_{1}^{(\lambda )}$ are subjected to boundary conditions $\tilde{{\cal P}}%
_{0}^{(\lambda )}(a)=\tilde{{\cal P}}_{0}^{(\lambda )}(-a),\tilde{{\cal P}}%
_{1}^{(\lambda )}(a)=\tilde{{\cal P}}_{1}^{(\lambda )}(-a)$. The
equation satisfied by the superfield monodromy operator
$T_{\lambda }(x,\theta )$ is
\begin{equation}
\frac{\partial }{\partial x}T_{\lambda }(x,\theta )=\tilde{{\cal P}}%
_{1}^{(\lambda )}T_{\lambda }(x,\theta ),  \label{mon}
\end{equation}
with the boundary condition $T_{\lambda }(-a)=1.$ The solution of
(\ref{mon}) is
\[
T_{\lambda }(x,\theta )=P\exp \left( -\int_{-a}^{x}dy\tilde{{\cal P}}%
_{1}^{(\lambda )}(y,\theta )\right) ,
\]
where $P$ is the path-ordered operator. The operator $T_{\lambda
}(x,\theta ) $ obeys
\[
\frac{\partial }{\partial t}T_{\lambda }(x,\theta )=\left[ \tilde{{\cal P}}%
_{0}^{(\lambda )},T_{\lambda }(a)\right] ,
\]
which is equivalent to the Lax formalism. This can be used to
generate an infinite sequence of local and non-local conservation
laws as detailed in the next sections.

\section{Superfield conserved quantities}\label{section4}

\subsection{Local conserved quantities}
We now derive the continuity equations of the superfield local
conserved quantities of the model via a set of superfield Riccati
equations using the method adopted for the bosonic models (see for
example \cite{sch,LEE}). The equation of motion (\ref{con2}) can
be written as
\begin{equation}
{\cal D}_{-}K_{+}-{\cal D}_{+}K_{-}\equiv D_{-}K_{+}+i\left\{ {\cal A}%
_{-},K_{+}\right\} -D_{+}K_{-}-i\left\{ {\cal A}_{+},K_{-}\right\}
=0, \label{lc1}
\end{equation}
where
\[
K_{\pm }\equiv {\cal J}_{\pm }-{\cal A}_{\pm }=-iG^{-1}D_{\pm }G-{\cal A}%
_{\pm }.
\]
The equation (\ref{lc1}) immediately  gives
\begin{equation}
2{\cal D}_{-}K_{+}+2i\left\{ {\cal A}_{-},K_{+}\right\} =-i\left\{
K_{+},K_{-}\right\} -{\cal F}_{\,-+}.  \label{lc2}
\end{equation}
Since ${\cal G}/{\cal H}$ is a symmetric space, therefore by using
equation (\ref{symmetric}) the left- and right-hand sides of
equation (\ref {lc2}) must vanish separately, i.e.
\begin{equation}
{\cal D}_{-}K_{+}=-i\left\{ {\cal A}_{-},K_{+}\right\} \,\,\,\,\,\,\mbox{in}%
\,\,\,{\bf k}\,,\,\,\,\,\,\,\,\,\,\,\,\,\,\,\,\,\,\,\,\,\,\,\,\,\,\,\,\,\,\,%
\,{\cal F}_{\,-+}=-i\left\{ K_{+},K_{-}\right\} \,\,\,\,\,\mbox{in}%
\,\,\,\,\,\,\,\,{\bf h.}  \label{lc3}
\end{equation}
These considerations lead to essentially two classes of local
conserved quantities; one class consists of currents based on
generators of the de Rham cohomology ring of ${\cal G}/{\cal H}$,
and second class consists of currents that are higher-spin
generalization of the super energy momentum tensor \cite{J3}.

An infinite series of local conservation laws can also be obtained
by using a set of compatible Riccati equation. A set of compatible
Riccati equations for the symmetric space sigma model in
superspace is given by
\begin{eqnarray}
D_{+}{\cal N}(\gamma ) &=&-i\gamma ^{-1}K_{+}-i\gamma ^{-1}{\cal
N}(\gamma
)K_{+}{\cal N}(\gamma )+i\left[ {\cal N}(\gamma ),\cal{A}_{+}\right] ,  \nonumber \\
D_{-}{\cal N}(\gamma ) &=&i\gamma K_{-}+i\gamma {\cal N}(\gamma )K_{-}{\cal %
N}(\gamma )+i\left[ {\cal N}(\gamma ),\cal{A}_{-}\right] ,
\label{sr1}
\end{eqnarray}
where ${\cal N}\in {\cal G}/{\cal H}$ is an even matrix
superfield. The following conservation equation immediately
follows
\begin{equation}
\gamma ^{-1}D_{-}\mbox{Tr}\left( {\cal N}(\gamma )K_{+}\right) -\gamma D_{+}%
\mbox{Tr}\left( {\cal N}(\gamma )K_{-}\right) =0.  \label{sr2}
\end{equation}
An infinite sequence of local conservation laws can be obtained by
expanding
${\cal N}(\gamma )$ as power series in $\gamma :{\cal N}(\gamma )=%
\sum_{k=0}^{\infty }\gamma ^{k}{\cal N}_{k}$. On substituting this
expansion in equation (\ref{sr2}), one arrives at the algebraic
equations obtain successively. The coefficients can be determined
by these algebraic equations and substitution of these
coefficients yields explicit expressions of the conserved
quantities. The details of this depends on the particular model.

The existence of local conserved quantities in supersymmetric
models on symmetric spaces can have certain relations with the
integrable structures in superstring theory on $AdS_{5}\times
S^{5}$ where the theory has been regarded as a non-linear sigma
model with the field taking values in the supercoset space
\begin{eqnarray}
\frac{PSU(2,2\mid4)}{SO(4,1)\times SO(5)}.\label{ads1}
\end{eqnarray}
The even part of this space is
\begin{eqnarray}
\frac{SO(4,2)}{SO(4,1)}\times \frac{SO(6)}{SO(5)}=AdS_{5}\times
S^{5}\nonumber
\end{eqnarray}
which is a symmetric space and therefore admits a Lax formalism
(one-parameter family of flat connections) and can further be
related to the conserved quantities on the Yang-Mills sector of
the $AdS/CFT$ correspondence \cite{ads1}-\cite{ads8}. It is also
worthwhile to study how this coincides with the local conserved
quantities for the models based on supercoset spaces. In
\cite{ads9} it has been shown that the classical superstring
theory on $AdS_{5}\times S^{5}$ as a supercoset sigma model,
admits a Lax formalism which does not imply the existence of local
conserved quantities. In the light of our results, we expect that
the existence of local conserved quantities might appear from the
Lax forlamism of the model on the even part of the supercoset
space which defines a symmetric space. In all these investigations
of integrable structures of classical superstring theory, a target
space supersymmetry has been used while the supersymmetric models
we have studied involve world-sheet supersymmetry. At this stage
we have not been able to relate the integrable structures of these
supersymmetric theories. The formalism we have developed can be
extended to the model with target space supersymmetry.

\subsection{Non-local conserved quantities}
The non-local conserved quantities have been constructed for both
the bosonic as well as the supersymmetric sigma model, via a
family of flat currents \cite{Pohl}-\cite{hassan}. For our model,
we assume spatial boundary conditions such that the superfields
$J_{\pm }$ vanish as $x\rightarrow \pm \infty $. The equation
(\ref{even2}), then implies that ${\cal U}(t,\pm \infty ,\theta
;\lambda )$ are independent of time. The residual freedom in the
solution for {${\cal U}$}${^{(\gamma )}}$ allows us to fix ${\cal
U}(t,-\infty ,\theta ;\lambda )$ equal to a unit matrix. We are
then left with a time independent function, ${\cal Q}(\lambda
)\,=\,{\cal U}(t,\infty ;\lambda )$. Expanding ${\cal Q}(\lambda
)$ as a power series in $\lambda $ gives infinitely many conserved
quantities
\[
{\cal Q}({\lambda })\,=\,\sum_{k=0}^{\infty }\lambda ^{k}{\cal Q}%
^{(k)}\;,\qquad {\frac{d{{\cal Q}^{(k)}}}{\ dt}}\,=\,0.
\]
In order to derive explicit expressions for these conserved
quantities in terms of superfields, we write equation
(\ref{even2}) as
\begin{eqnarray}
{\cal U}(t,x,\theta ;\lambda ) &=&1\,+\frac{1}{2}\int_{-\infty
}^{x}dy{\cal U}(t,y,\theta ;\lambda )\left\{ \left( \frac{2\lambda
}{1-\lambda }\right) D_{+}J_{+}+\left(
\frac{2\lambda }{1+\lambda }\right) D_{-}J_{-}\right.  \nonumber \\
&&\,\,\,\left. -i\left( \frac{2\lambda }{1-\lambda }\right)
^{2}J_{+}^{2}+i\left( \,\frac{2\lambda }{1+\lambda }\right)
^{2}J_{-}^{2}\;\right\}. \label{sPCM18}
\end{eqnarray}
We expand the superfield ${\cal U}(t,x,\theta ;\lambda )$ as a
power series in $\lambda $,
\begin{equation}
{\cal U}(t,x,\theta ;\lambda )=\sum_{k=0}^{\infty }\lambda ^{k}{\cal U}%
_{k}(t,x,\theta ),  \label{exp}
\end{equation}
and compare the coefficients of powers of $\lambda $, one gets a
series of conserved non-local superfield currents, which upon
integration give non-local conserved quantities. The expressions
for the first few cases are
\begin{eqnarray*}
{\cal Q}^{(1)a} &=&\int_{-\infty }^{\infty }dy\,\,\left(
\,D_{+}J_{+}^{a}+D_{-}J_{-}^{a}\,\right) (t,y,\theta )\;, \\
{\cal Q}^{(2)a} &=&\int_{-\infty }^{\infty
}dy\left((\,D_{+}J_{+}^{a}-D_{-}J_{-}^{a}\,)(t,y,\theta
)-if\,^{abc}(J_{+}^{b}J_{+}^{c}-J_{-}^{b}J_{-}^{c}\,)(t,y,\theta )\right. \\
&&\left.
+\frac{1}{2}f\,^{abc}(D_{+}J_{+}^{b}+D_{-}J_{-}^{b}\,)(t,y,\theta
)\,\int_{-\infty
}^{y}dz\,\,(D_{+}J_{+}^{c}+D_{-}J_{-}^{c}\,)(t,z,\theta )\,\right)
\;.
\end{eqnarray*}
These are the desired non-local conserved quantities which
corresponds to the bosonic non-local conserved quantities of
\cite{EIC} when the fermions are set to zero. The component
content of these superfield conserved quantities is the same which
appeared for certain models in \cite{Curt}-\cite{Cos}. The
non-local conserved quantities are also known to exist in the
classical theory of Green-Schwarz superstrings, where a parameter
dependent flat current taking values in Lie algebra of
$PSU(2,2|4)$ is shown to exist\cite{ads1}. The construction of
non-local conserved quantities has been extended to the case of
full supercoset space (\ref{ads1}) which is not symmetric and the
theory also involves a Wess-Zumino term and $\kappa$-symmetry
\cite{Val,ber}. Moreover, it has been shown that Yangian non-local
symmetries exist in $D=4$ superconformal Yang-Mills theory in the
gauge theory sector of the $AdS/CFT$ correspondence \cite{dolan}.
\section{Examples}\label{section5}

\subsection{Supersymmetric model on complex Grassmannian}

In the previous section, we have discussed a general procedure of
studying the Lax formalism and extracting conserved quantities for
a symmetric space sigma model in superspace. In this section we
will discuss an example i.e.
the sigma model on the complex Grassmannian manifold $U(m+n)/U(m)\times U(n)$%
. For $n=1$, it reduces to the complex projective space $CP^{m}$.
We define
a $U(m+n)$ valued matrix superfield $G(x^{\pm },\theta ^{\pm })$%
\[
G(x^{\pm },\theta ^{\pm })=\left(
\begin{array}{l}
X \\
Y
\end{array}
\right) ,\;\;\;\;\;\;\;\;\;G^{\dagger }(x^{\pm },\theta ^{\pm
})G(x^{\pm },\theta ^{\pm })=I=G(x^{\pm },\theta ^{\pm
})G^{\dagger }(x^{\pm },\theta ^{\pm }),
\]
where $X(x^{\pm },\theta ^{\pm })$ and $Y(x^{\pm },\theta ^{\pm
})$ are superfield matrices of order $m\times (m+n)$ and $n\times
(m+n)$ respectively. We introduce orthogonal projectors for
superfields
\[
P=XX^{\dagger },\,\,\,\,\,\,\,\,\,\,\bar{P}=YY^{\dagger
},\,\,\,\,\,\,\,\,\,\,\,\,\,\,\,\,\,\,\,\,P+\bar{P}=I,
\]
which map $C^{m+n}$ in to $m$ and $n$ dimensional subspaces
spanned by the column vectors of $X$ and $Y$ respectively. The
super gauge transformation acts on superfield $G(x^{\pm },\theta
^{\pm })$ as
\[
G(x^{\pm },\theta ^{\pm })=\left(
\begin{array}{l}
X \\
Y
\end{array}
\right) \rightarrow G^{\prime }(x^{\pm },\theta ^{\pm })=\left(
\begin{array}{ll}
H_{1} & 0 \\
0 & H_{2}
\end{array}
\right) \left(
\begin{array}{l}
X \\
Y
\end{array}
\right) ,
\]
where
\[
\left(
\begin{array}{ll}
H_{1} & 0 \\
0 & H_{2}
\end{array}
\right) \in U(m)\times U(n).
\]
The canonical decomposition of the superfield $D_{\pm }GG^{-1}$ is
given by
\begin{eqnarray}
{\cal A}_{\pm } &=&\left(
\begin{array}{ll}
iD_{\pm }X\,X^{\dagger } & 0 \\
0 & iD_{\pm }YY^{\dagger }
\end{array}
\right) ,  \nonumber \\
K_{\pm } &=&\left(
\begin{array}{ll}
0 & iD_{\pm }XY^{\dagger } \\
iD_{\pm }YX^{\dagger } & 0
\end{array}
\right) .  \label{sr3}
\end{eqnarray}
The even superfield matrix ${\cal N}$ appearing in equations (\ref{sr1})-(%
\ref{sr2}), decomposes as
\begin{equation}
{\cal N=}\left(
\begin{array}{ll}
0 & -{\cal M}^{\dagger } \\
{\cal M} & 0
\end{array}
\right) ,  \label{sr4}
\end{equation}
where ${\cal M}$ is an $n\times m$ even matrix superfield. The
action of covariant derivative in superspace on the superfield
$G(x^{\pm },\theta ^{\pm })$ will be
\[
{\cal D}_{\pm }\left(
\begin{array}{l}
X \\
Y
\end{array}
\right) \equiv \left(
\begin{array}{l}
{\cal D}_{\pm }X \\
{\cal D}_{\pm }Y
\end{array}
\right) =\left(
\begin{array}{l}
i\bar{P}D_{\pm }X \\
iPD_{\pm }Y
\end{array}
\right) .
\]
As a result of decompositions, the action for the complex
Grassmannian model splits into two parts given as
\[
{\cal L}\equiv {\frac{1}{2}}\int d^{2}xd^{2}\theta \mbox{Tr}({\cal D}%
_{+}G^{-1}{\cal D}_{-}G)={\frac{1}{2}}\int d^{2}xd^{2}\theta
\left(
\mbox{Tr}\left( {\cal D}_{+}X({\cal D}_{-}X)^{\dagger }\right) +\mbox{Tr}%
\left( {\cal D}_{+}Y({\cal D}_{-}Y)^{\dagger }\right) \right) .
\]
The variation of the action yields following equations of motion
for each superfields $X$ and $Y$
\begin{equation}
{\cal D}_{+}{\cal D}_{-}X-X({\cal D}_{+}X)^{\dagger }{\cal D}%
_{-}X=0,\,\,\,\,\,\,{\cal \,D}_{+}{\cal D}_{-}Y-Y({\cal D}_{+}Y)^{\dagger }%
{\cal D}_{-}Y=0.\,\,\,\,\,\,\,\,\,\,\,\,\,\,\,\,  \label{d1}
\end{equation}
We rewrite the above equations of motion in terms of projector
superfields
\[
\left[ D_{+}D_{-}P,P\right] =0=\left[
D_{+}D_{-}\bar{P},\bar{P}\right] .
\]
The one-parameter family of transformations acts on superfields
$X$ and $Y$ as
\[
\left(
\begin{array}{l}
X \\
Y
\end{array}
\right) \rightarrow {\cal U}^{(\gamma )}\left(
\begin{array}{l}
X \\
Y
\end{array}
\right) ,
\]
and the corresponding linear system can be expressed as
\begin{eqnarray*}
D_{+}{\cal U}^{(\gamma )} &\equiv &-i(1-\gamma ^{-1}){\cal U}^{(\gamma )}\,%
{\cal D}_{+}\left(
\begin{array}{l}
X \\
Y
\end{array}
\right) \left(
\begin{array}{l}
X \\
Y
\end{array}
\right) ^{\dagger }, \\
D_{-}{\cal U}^{(\gamma )} &\equiv &-i(1-\gamma ){\cal U}^{(\gamma )}\,{\cal D%
}_{-}\left(
\begin{array}{l}
X \\
Y
\end{array}
\right) \left(
\begin{array}{l}
X \\
Y
\end{array}
\right) ^{\dagger },
\end{eqnarray*}
where
\[
-{\cal D}_{\pm }\left(
\begin{array}{l}
X \\
Y
\end{array}
\right) \left(
\begin{array}{l}
X \\
Y
\end{array}
\right) ^{\dagger }=\left[ D_{\pm }P,P\right] =\left[ D_{\pm }\bar{P},\bar{P}%
\right] .
\]
The projector superfields $P$ and $\bar{P}$ transform according to
the law
\[
P\rightarrow {\cal U}^{(\gamma )}P{\cal U}^{(\gamma )\dagger
},\,\,\,\,\,\,\,\,\,\,\,\,\,\,\,\,\,\,\,\,\bar{P}\rightarrow {\cal U}%
^{(\gamma )}\bar{P}{\cal U}^{(\gamma )\dagger }.
\]
One can apply the same procedure to the real Grassmannian
manifold. These considerations are sufficient for the construction
of conserved quantities of the model. The manifold is a symmetric
space when we define an involutive automorphism $\sigma $ acting
on the superfield $G$ as
\[
\sigma (G)=\Theta G\Theta ^{-1},
\]
where
\[
G(x^{\pm },\theta ^{\pm })\in
U(m+n)\,,\,\,\,\,\,\,\,\,\,\,\,\,\,\,\,\,\,\,\,\,\,\,\,\,\,\,\,\,\,\,\,\,\,%
\,\,\,\,\,\,\,\,\,\,\,\,\Theta =\left(
\begin{array}{ll}
I_{m} & 0 \\
0 & -I_{n}
\end{array}
\right) .
\]
Using equations (\ref{sr3})-(\ref{sr4}), the equations (\ref{sr1})-(\ref{sr2}%
) generate an infinite series of local conservation laws in term
of superfields $B_{\pm }=iD_{\pm }X\,X^{\dagger }$, $C_{\pm
}=iD_{\pm }YY^{\dagger }$and $K_{\pm }$. The set of compatible
Riccati differential equations for an even matrix superfield
${\cal M}$ is
\begin{eqnarray*}
D_{+}{\cal M}(\gamma ) &=&-i\gamma ^{-1}K_{+}+i\gamma ^{-1}{\cal
M}(\gamma )K_{+}^{\dagger }{\cal M}(\gamma )-iC_{+}{\cal M}(\gamma
)+i{\cal M}(\gamma
)B_{+}, \\
D_{-}{\cal M}(\gamma ) &=&i\gamma K_{-}-i\gamma {\cal M}(\gamma)K_{-}^{\dagger }%
{\cal M}(\gamma)-iC_{-}{\cal M}(\gamma )+i{\cal M}(\gamma )B_{-}.
\end{eqnarray*}
This set can immediately be used to derive an infinite sequence of
conservation laws
\[
\gamma ^{-1}D_{-}\mbox{Tr}\left( {\cal M}^{\dagger }(\gamma )K_{+}+{\cal M}%
(\gamma )K_{+}^{\dagger }\right) -\gamma D_{+}\mbox{Tr}\left( {\cal M}%
^{\dagger }(\gamma )K_{-}+{\cal M}(\gamma )K_{-}^{\dagger }\right)
=0.
\]
Expanding ${\cal M}(\gamma )$ as power series in $\gamma :{\cal M}(\gamma )=%
\sum_{k=0}^{\infty }\gamma ^{k}{\cal M}_{k}$, one can generate
$\gamma $-independent conservation laws.
\subsection{Supersymmetric principal chiral model (SPCM)}
In this section, we discuss the supersymmetric principal chiral
model (SPCM) as a symmetric space model. If we suppose ${\cal H}$
is trivial subgroup of $\cal G$, setting ${\cal A}_{\pm }=0$, and
$K_{\pm }$ becomes
\[
K_{\pm }\rightarrow -iG^{-1}D_{\pm }G={\cal J}_{\pm }.
\]
Equation (\ref{lc2}) becomes
\begin{equation}
D_{-}{\cal J}_{+}=-\frac{i}{2}\left\{ {\cal J}_{+},{\cal
J}_{-}\right\} . \label{EOM1}
\end{equation}
The Lie group ${\cal G}$ can now be considered as a symmetric
space: let
\[
\Delta {\cal G}=\left\{ (G,G)\left| G\in {\cal G}\right. \right\}
,
\]
be the diagonal of ${\cal G}\times {\cal G}$, and define $\sigma :{\cal G}%
\times {\cal G}\rightarrow {\cal G}\times {\cal G}$ such that
\[
\sigma (G,G^{\prime })=(G^{\prime },G).
\]
Then ${\cal G\times G}/\Delta {\cal G}$ is a symmetric space with
involution $\sigma .$ We define a map ${\cal G}\times {\cal G}$
$\rightarrow {\cal G}$ such that the pair $(G_{1},G_{2})$ is
mapped to $G=G_{1}G_{2}^{-1}$. In this case the decomposition of
the corresponding Lie algebra will be
\[
{\bf g}+{\bf g}={\bf h}+{\bf k}.
\]
By writing the superfield $G=(G_{1},G_{2})$ taking values in $%
{\cal G}\times {\cal G}$, the gauge superfield can be expressed as
\[
{\cal A}_{\pm }=\left( \left( -\frac{i}{2}G_{1}^{-1}D_{\pm}G_{1}-\frac{i}{2}%
G_{2}^{-1}D_{\pm}G_{2}\right) ,\left( -\frac{i}{2}G_{1}^{-1}D_{\pm}G_{1}-\frac{i%
}{2}G_{2}^{-1}D_{\pm}G_{2}\right) \right) .
\]
The one-parameter family of transformations on these fields is
given by
\[
(G_{1},G_{2})\rightarrow (G_{1}^{(\gamma )},G_{2}^{(\gamma )})=({\cal U}%
_{1}^{(\gamma )}G_{1},{\cal U}_{2}^{(\gamma )}G_{2})=({\cal
U}_{1}^{(\gamma )},{\cal U}_{2}^{(\gamma )})(G_{1},G_{2}),
\]
where {${\cal U}_{1}$}${^{(\gamma )}}$ and {${\cal
U}_{2}$}${^{(\gamma )}}$ belong to {${\cal G}$}. The
transformation on superfield $G(x^{\pm },\theta ^{\pm })$ is
therefore
\begin{equation}
G(x^{\pm },\theta ^{\pm })\mapsto G^{(\gamma )}(x^{\pm },\theta ^{\pm })\,=%
{\cal U}_{1}{^{(\gamma )}}G(x^{\pm },\theta ^{\pm }){\cal
U}_{2}{^{(\gamma )-1}}\;,  \label{dual}
\end{equation}
Here we choose the boundary values ${{\cal U}_{1}^{(1)}}=1$, $\,{{\cal U}%
_{2}^{(1)}}=1\,$or $G^{(1)}=G$. The set of linear differential
equations
satisfied by {${\cal U}_{1}$}${^{(\gamma )}}$ and {${\cal U}_{2}$}${%
^{(\gamma )}}$ are
\begin{eqnarray}
\left( D_{+}{\cal U}_{1}^{(\gamma )},D_{+}{\cal U}_{2}^{(\gamma
)}\right) &=&-(1-\gamma ^{-1})\left( {\cal U}_{1}^{(\gamma
)},{\cal U}_{2}^{(\gamma
)}\right) {\cal D}_{+}(G_{1},G_{2})(G_{1},G_{2})^{-1},  \nonumber \\
\left( D_{-}{\cal U}_{1}^{(\gamma )},D_{-}{\cal U}_{2}^{(\gamma
)}\right) &=&-(1-\gamma )\left( {\cal U}_{1}^{(\gamma )},{\cal
U}_{2}^{(\gamma )}\right) {\cal
D}_{-}(G_{1},G_{2})(G_{1},G_{2})^{-1}.  \label{Red1}
\end{eqnarray}
Evaluating the covariant derivative and using $G=G_{1}G_{2}^{-1},$
we arrive at
\begin{eqnarray*}
\left( D_{+}{\cal U}_{1}^{(\gamma )},D_{+}{\cal U}_{2}^{(\gamma
)}\right)
&=&i(1-\gamma ^{-1})\left( i{\cal U}_{1}^{(\gamma )}D_{+}GG^{-1},-i{\cal U}%
_{2}^{(\gamma )}G^{-1}D_{+}G\right) , \\
\left( D_{-}{\cal U}_{1}^{(\gamma )},D_{-}{\cal U}_{2}^{(\gamma
)}\right)
&=&i(1-\gamma )\left( i{\cal U}_{1}^{(\gamma )}D_{-}GG^{-1},-i{\cal U}%
_{2}^{(\gamma )}G^{-1}D_{-}G\right) .
\end{eqnarray*}
If we take ${\cal U}_{1}{^{(\gamma )}=}{\cal U}{^{(\gamma )}}$ and ${\cal U}%
_{2}{^{(\gamma )}=}{\cal V}{^{(\gamma )}}$ , the equations (\ref{dual}) and (%
\ref{Red1}) reduce to following equations
\begin{equation}
G(x^{\pm },\theta ^{\pm })\mapsto G^{(\gamma )}(x^{\pm },\theta ^{\pm })\,=%
{\cal U}{^{(\gamma )}}G(x^{\pm },\theta ^{\pm }){\cal V}{^{(\gamma
)-1}}\;, \label{dual1}
\end{equation}
\begin{eqnarray}
D_{+}{\cal U}{^{(\gamma )}} &=&{\frac{i}{2}}(1-\gamma ^{-1}){\cal U}{%
^{(\gamma )}}{\cal J}{_{+}^{L},}  \label{l1} \\
D_{-}{\cal U}{^{(\gamma )}} &=&{\frac{i}{2}}(1-\gamma ){\cal U}{^{(\gamma )}%
}{\cal J}_{-}^{L}{,}  \label{l2} \\
\;D_{+}{\cal V}{^{(\gamma )}} &=&{\frac{i}{2}}(1-\gamma ^{-1}){\cal V}{%
^{(\gamma )}}{\cal J}_{+}^{R},  \label{l3} \\
D_{-}{\cal V}{^{(\gamma )}} &=&{\frac{i}{2}}(1-\gamma ){\cal V}{^{(\gamma )}%
}{\cal J}_{-}^{R}\;.  \label{sPCM13}
\end{eqnarray}
where ${\cal J}{_{\pm }^{L}=}iD_{\pm }GG^{-1}$ and ${\cal J}_{\pm }^{R}{=-}%
iG^{-1}D_{\pm }G.$ The compatibility conditions for these
equations are
obtained by applying $D_{-}$ to the equations (\ref{l1}) and (\ref{l3}) and $%
D_{+}$ to equations (\ref{l2}) and (\ref{sPCM13}), so that one
gets
\begin{eqnarray*}
 {\cal U}^{(\gamma )}\left\{ (1-\gamma ^{-1})D_{-}{\cal J}_{+}^{L}+(1-\gamma )D_{+}{\cal J%
}_{-}^{L}+i(1-{{\frac{1}{2}}}(\gamma +\gamma ^{-1}))\{{\cal J}_{+}^{L}\,,\,%
{\cal J}_{-}^{L}\}\right\} &=&\,0, \\
{\cal V}^{(\gamma )} \left\{ (1-\gamma ^{-1})D_{-}{\cal J}_{+}^{R}+(1-\gamma )D_{+}{\cal J}%
_{-}^{R}+i(1-{{\frac{1}{2}}}(\gamma +\gamma ^{-1}))\{{\cal J}_{+}^{R}\,,\,%
{\cal J}_{-}^{R}\}\right\} &=&\,0.
\end{eqnarray*}
We see that the supersymmetric principal chiral model is in fact
an integrable supersymmetric sigma model on a symmetric space and
can easily be extracted from a supersymmetric sigma model on a
general symmetric space.

Let us write an arbitrary element of ${\cal G\times G}$ in the
form
\[
\Gamma =\left(
\begin{array}{ll}
G_{L} & 0 \\
0 & G_{R}
\end{array}
\right) ,
\]
and consider the symmetric space sigma model that corresponds to
the involutive automorphism given by
\[
\Sigma =\left(
\begin{array}{ll}
0 & 1 \\
1 & 0
\end{array}
\right) .
\]
We can now construct the superfields
\[
\tilde{\Gamma}\,\equiv \Sigma G^{-1}\Sigma=\left(
\begin{array}{ll}
G_{R}^{-1} & 0 \\
0 & G_{L}^{-1}
\end{array}
\right),
\]
and
\[
M\equiv \tilde{\Gamma}\Gamma =\left(
\begin{array}{ll}
G_{R}^{-1}G_{L} & 0 \\
0 & G_{L}^{-1}G_{R}
\end{array}
\right) .
\]
The subgroup consists of diagonal elements for which
$G_{R}=G_{L}$, and write $G=G_{R}^{-1}G_{L},$ so that
\[
M=\left(
\begin{array}{ll}
G & 0 \\
0 & G^{-1}
\end{array}
\right) .
\]
Then the superfield conserved currents of the model can be written
as
\[
{\cal J}_{\pm }\equiv iM^{-1}D_{\pm }M=\left(
\begin{array}{ll}
iG^{-1}D_{\pm }G & \,\,\,\,\,\,\,\,\,0 \\
\,\,\,\,\,\,\,\,\,\,\,\,\,\,0 & iGD_{\pm }^{-1}G
\end{array}
\right).
\]
The construction then yields the superfield Lax formalism of
supersymmetric principal chiral model (SPCM). The Lax formalism of
SPCM is responsible for the existence of an infinite sequence of
local and non-local conserved quantities \cite{hassan}.

The superspace equation of motion (\ref{EOM1}) implies an infinite
series of local conservation laws \cite{EHMM}-\cite{hassan}
\[
D_{\pm }\mbox{Tr}\left( {\cal J}_{\mp }\right) ^{m}=0,\,\,\,D_{\pm }\mbox{Tr}%
\left( {\cal J}_{\mp }^{m-1}{\cal J}_{\mp \mp }\right)=0 ,\,\,\,\mbox{with }%
{\cal J}_{\mp \mp }=D_{\mp }{\cal J}_{\mp }+i{\cal J}_{\mp }^{2},
\]
where the values of $m$ are precisely the exponents of the Lie algebra of $%
{\cal G}$. The local conserved quantities of SPCM also arise from
the Lax formalism via super B\"{a}cklund transformation (SBT) or
equivalently from super Riccati equations \cite{hassan}. One can
easily obtain an infinite sequence of non-local conserved
quantities for the SPCM. The expressions for the first two
non-local conserved quantities are
\begin{eqnarray*}
\tilde{Q} ^{(1)a} &=&\frac{1}{2}\int_{-\infty }^{\infty }dy\,\,(\,D_{+}{\cal J}%
_{+}^{a}+D_{-}\,{\cal J}_{-}^{a})(t,y,\theta )\;, \\
\tilde{Q}^{(2)a} &=&\int_{-\infty }^{\infty }dy\,\,\left(\frac{1}{2}%
(\,D_{+}{\cal J}_{+}^{a}-D_{-}{\cal J}_{-}^{a}\,)(t,y,\theta )-\frac{i}{4}%
f\,^{abc}({\cal J}_{+}^{b}{\cal J}_{+}^{c}-{\cal J}_{-}^{b}{\cal J}%
_{-}^{c}\,)(t,y,\theta )\right. \\
&&\left. +\,\frac{1}{8}f\,^{abc}(D_{+}{\cal J}_{+}^{b}+D_{-}{\cal J}%
_{-}^{b}\,)(t,y,\theta )\,\int_{-\infty }^{y}dz\,\,(D_{+}{\cal J}%
_{+}^{c}+D_{-}{\cal J}_{-}^{c}\,)(t,z,\theta )\,\right) \;.
\end{eqnarray*}
These conserved quantities are exactly the same as obtained in
\cite{hassan}. The component contents gives bosonic conserved
quantities which generates a Yangian with two copies corresponding
to left and right currents \cite{Cos}-\cite{hassan}.
\section{Concluding remarks}\label{section6}

We have investigated a one-parameter family of flat superfield
connections of supersymmetric sigma model based on symmetric
spaces. This suggests that the model in superspace represents an
integrable system exhibiting Lax formalism and the existence of an
infinite number of local and non-local conserved quantities. Some
explicit examples are given to illustrate the results. The work
can be extended to a number of directions. The immediate study
that needs a consideration is the quantization of the conserved
quantities which could eventually lead to the implications of the
$S$-matrices of these models. It will be interesting to develop a
similar formalism for the superspace sigma models based on
supercoset spaces which appear in the superstring theory on the
$AdS_{5}\times S^{5}$ background and its relation to Yangian
symmetry. The $r$-matrix formalism of the superspace sigma models
based on bosonic symmetric spaces and supercoset spaces is also a
direction which needs to be investigated. Moreover many of the
integrability structures which have appeared in $AdS/CFT$
correspondence can be further extended to incorporate certain
mathematical techniques of the integrable field theories such as
involution of local conserved quantities, their quantization,
$S$-matrix and $r$-matrix formalism, algebraic and thermodynamics
Bethe Ansatz etc.

{\Large Acknowledgements}

The authors acknowledge the enabling role of the Higher Education
Commission Islamabad, Pakistan and appreciate its financial
support through ``Merit Scholarship Scheme for PhD studies in
Science \& Technology (200 Scholarships)''.

\end{document}